\documentclass[10pt]{iopart}

\usepackage{supertabular}
\usepackage{graphicx}
\usepackage{xspace}
\usepackage{hyperref}
\usepackage{orcidlink}
\usepackage{iopams}  
\usepackage{natbib}

\begin{document}

\title[In-flight pixel degradation of the Sentinel 5 Precursor TROPOMI-SWIR HgCdTe detector]{In-flight pixel degradation of the Sentinel 5 Precursor TROPOMI-SWIR HgCdTe detector}

\author{Tim A. van Kempen$^1$\orcidlink{0000-0001-8178-5054}, Marina Lobanova$^2$, Richard van Hees$^1$\orcidlink{0000-0002-3846-0753}, Valentina Masarotto$^2$\orcidlink{0000-0002-2133-7660}, Paul Tol$^1$, Solomii Kurchaba$^3,1$\orcidlink{0000-0002-0202-1898}, Ruud W. M. Hoogeveen$^1$\orcidlink{0000-0001-7346-0276}}

\address{$^1$ SRON Netherlands Institute for Space Research, Niels Bohrweg 4,  2333 CA, Leiden, the Netherlands}
\address{$^2$ Mathematisch Instituut, Leiden University, Faculty of Science, Niels Bohrweg 1, 2333 CA, Leiden, the Netherlands}
\address{$^3$ Leiden Institute of Advanced Computer Science, Faculty of Science, Niels Bohrweg 1, 2333 CA, Leiden, the Netherlands}
\ead{t.a.van.kempen@sron.nl}
\vspace{10pt}
\begin{indented}
\item[]February 2024
\end{indented}
\begin{abstract}
The TROPOMI-SWIR HgCdTe detector on the Sentinel-5 Precursor mission has been performing in-orbit measurements of molecular absorption in Earth's atmosphere since its launch in October 2017. In its polar orbit the detector is continuously exposed to potentially harmful energetic particles. Calibration measurements taken during the eclipse are used to inspect the performance of this detector. This paper explores the in-orbit degradation of the HgCdTe detector.
After five years, the detector is still performing within specifications, even though pixels are continuously hit by cosmic radiation. The bulk of the impacts have no lasting effects, and most of the damaged pixels (95$\%$) appear to recover on the order of a few days to several months, attributed to a slow spontaneous recovery of defects in the HgCdTe detector material. This is observed at the operational temperature of 140 K.  The distribution of the observed recovery times has a mean around nine days with a significant tail towards several months. Pixels that have degraded have a significant probability to degrade again. The location of faulty pixels follows a Poissonian distribution across the detector. No new clusters have appeared, revealing that impacts are dominated by relatively low energetic protons and electrons. Due to the observed spontaneous recovery of pixels, the fraction of pixels meeting all quality requirements in the nominal operations phase has always been over 98.7$\%$. 
The observed performance of the TROPOMI-SWIR detector in-flight impacts selection criteria of HgCdTe detectors for future space instrumentation.  

\end{abstract}

%
\vspace{2pc}
%
\submitto{\MST}
%
\maketitle
%
\ioptwocol

\section{Introduction}\label{sec:Introduction}
Launched on October 13th 2017, the Sentinel-5 precursor (S5p) mission has a single payload: the Tropospheric Monitoring Instrument, or TROPOMI\footnote{TROPOMI is a collaboration between Airbus Defence and Space Netherlands, KNMI, SRON and TNO, on behalf of NSO and ESA. Airbus Defence and Space Netherlands is the main contractor for the design, building and testing of the instrument. KNMI and SRON are the principal investigator institutes for the instrument. TROPOMI is funded by the following ministries of the Dutch government: the Ministry of Economic Affairs, the Ministry of Education, Culture and Science, and the Ministry of Infrastructure and the Environment.} \citep{Veefkind12}. 
TROPOMI consists of two modules: a) the UVN module covering ultraviolet (UV), visible (VIS) and near-infrared (NIR) wavelengths and b) the SWIR module, covering short-wavelength infrared (SWIR) wavelengths between 2305 and 2385 nm. The measured radiances of the SWIR module are used to determine the column densities of CO, H$_2$O and CH$_4$ \citep[e.g.,][]{Borsdorff18,Hu18}, and subsequently probe Earth's atmosphere composition \citep[e.g.][]{vanderVelde21,Maasakkers22,Schuit23}.

TROPOMI has made use of its significant improvements over prior satellites, such as OMI and SCIAMACHY, probing these species. It offers worldwide daily coverage using the push-broom concept and a large swath covering over 2500 km on the ground.  See e.g. \citet{Hubert20} for a comparison on tropospheric ozone. The simultaneous coverage of the UV, VIS, NIR and SWIR wavelengths also removes uncertainties in comparisons where previous generations of satellites struggled. The spatial resolution is also much higher than that of predecessors such as SCIAMACHY or OMI, thus producing much higher detail \citep[e.g., for methane see][]{Hu18}.

The SWIR module\footnote{developed by SSTL, United Kingdom} uses an immersed grating\footnote{developed by SRON \citep{Amerongen17}} as its dispersive element, while the Saturn SW detector was made by Sofradir, currently known as LYNRED. The HgCdTe\footnote{HgCdTe is also referred to as MCT. It consist of an alloy of Mercury Cadmium and Telluride. For consistency, we use HgCdTe} substrate offers a size of 1000 by 256 pixels \citep{Lei15}. Within the TROPOMI-SWIR module, the long edge of the detector is used to spectrally cover the 80 nm wide band, while the spatial direction is projected on the short edge, covering the projected 2500 km wide TROPOMI swath. The detector has a pixel pitch of 30 $\mu$m and its focal plane array is operated at 140 K. 

Before launch, the expected perfomance of the SWIR detector \citep{Hoogeveen13}, was verified through on-ground calibration  campaigns at TNO (Rijswijk, the Netherlands), SRON (Utrecht, the Netherlands) and CSL (Liege, Belgium). For specific results, we refer the reader to \citet{Tol18} on straylight performance, \citet{vanHees18} on the instrumental spectral response function and \citet{Kleipool18} for the full TROPOMI instrument.  The in-flight performance of TROPOMI has been, and continues to be, monitored using the on-board calibration unit, measurements on thermally controlled dark surfaces and TROPOMI's capability to daily measure the solar irradiance \citep[][as well as the SWIR monitoring website \url{http://www.sron.nl/tropomi-swir-monitoring}]{vanKempen19, Ludewig20, vanKempen23}. 

During nominal operations, the detector should only be illuminated by Earth's radiance, Solar irradiance over two diffusers, or various calibrations sources. Light from other wavelengths or sources should not be able to reach the detector due to the usage of a field stop and the immersed grating. Straylight (i.e. light that reaches the detector outside of the intended optical path) is suppressed during data processing following on-ground reference measurements \citep{Tol18}. The remaining largest non-instrumental source that induces charge on the detector is cosmic rays. At the orbit of S5p there are predominantly protons and electrons that populate the Van Allen radiation belts.
At the altitude of the orbit of S5p, these Van Allen radiation belts shield satellites from most, but not all, energetic particles by controlling their locations and movement through their interaction with Earth's magnetic field \citep{Koskinen22a, Koskinen22b}. As a consequence of the difference in position between Earth's dipole point and its center of mass, the belts are not symmetric \citep{Koskinen22a}. At the orbit altitude of TROPOMI (824 km), an area where a significantly larger amount of cosmic rays can reach the detector is the South Atlantic Anomaly (SAA). Here, the inner Van Allen radiation belt comes relatively close to the Earth's surface, below the S5p orbit. TROPOMI-SWIR is exposed to charged particles captured during passages through this region. Less pronounced enhancements occur near the Earth's poles. Higher energy particles (i.e., heavier ions), seen in locations beyond Earth \citep[e.g., at L2, see][]{Birkmann22, Rieke23}, are absent, having been captured by the outer Van Allen radiation belt \citep{Koskinen22a}. 

The effect of continuous impact of radiation on the performance of infrared detectors is not well-researched. \citet{Kimble08} desribes effects for the Hubble Space Telescope, and \citet{Gloudemans05}, \citet{Lichtenberg06}, \citet{Kleipool07} and \citet{Hilbig20} for SCIAMACHY. It is highly relevant for space-based instrumentation, as seen in the degradation of near- and mid-infrared detectors of the James Webb Space Telescope (JWST) \citep{Rauscher12,Birkmann22,Wright23,Rieke23}.

In this paper, we will present results on performance of the TROPOMI-SWIR detector during nearly six years in orbit. Section \ref{sec:detector} provides details on the Saturn detector, including the noise and dark current performance. The methodology used in quantifying this performance is given in section \ref{sec:methods} with results in section \ref{sec:inflight}. A statistical analysis is given in section \ref{sec:analysis}. Discussion on the results is given in section \ref{sec:discussion}. The final conclusions are listed in section \ref{sec:conclusions}.

\section{Detector}\label{sec:detector}
\subsection{Detector casing and protection}

The detector is part of the SWIR detector sub-assembly \citep{Hoogeveen13,Doornink17}. This includes connections to the front end electronics, the detector mounting plate, the detector cold finger as well as the interface plate. The mounting and interface plate are connected through a thermally isolating structure, to ensure thermal stability of the detector at 140 K. The mounting plate is made of molybdenum, a material chosen for high thermal conductivity. It also has the lowest thermal expansion coefficient of commercially available materials and provides significant shielding to the detector. The interface plate is made of a titanium alloy. The  sub-assembly features a warm window of germanium in front of the detector. It is fully Electromagnetic Compatible (EMC) shielded by the titanium alloy structure and the germanium warm window.
The entire SWIR instrument is shrouded by two aluminium alloy covers, designed to minimize stray light. The only opening passes through the immersed grating. The immersed grating prism is a single crystal of silicon with a grating surface etched onto one face, mounted on a titanium alloy structure. The active layer of the detector consists of a layer of HgCdTe on a lattice consisting of a CdZnTe substrate. The read-out integrated circuit (ROIC) performs in snapshot operation and can be read out while integrating. Although windowing is available, it is not used: the detector is read out completely at all times. The detector has a full-well capacity of 5 10$^5$ e$^-$, with an offset bias voltage near 0.55 V. The latter was a choice to improve performance.

Following the SCIAMACHY in-flight detector performance \citep{Gloudemans05,Lichtenberg06,Hilbig20}, major design drivers were adopted to the control of moisture and outgassing within the detector assembly and the prevention of the freezing of water on the detector window \citep[e.g.][ and the proceedings of ESA Contamination of Optical Equipment Workshop, December 2003]{Hoogeveen13}.  Before launch, extensive testing was carried out. See \citet{Hoogeveen13,Kleipool18,Tol18,vanHees18}. 

\begin{figure}[tp]
\includegraphics[width=8.5cm]{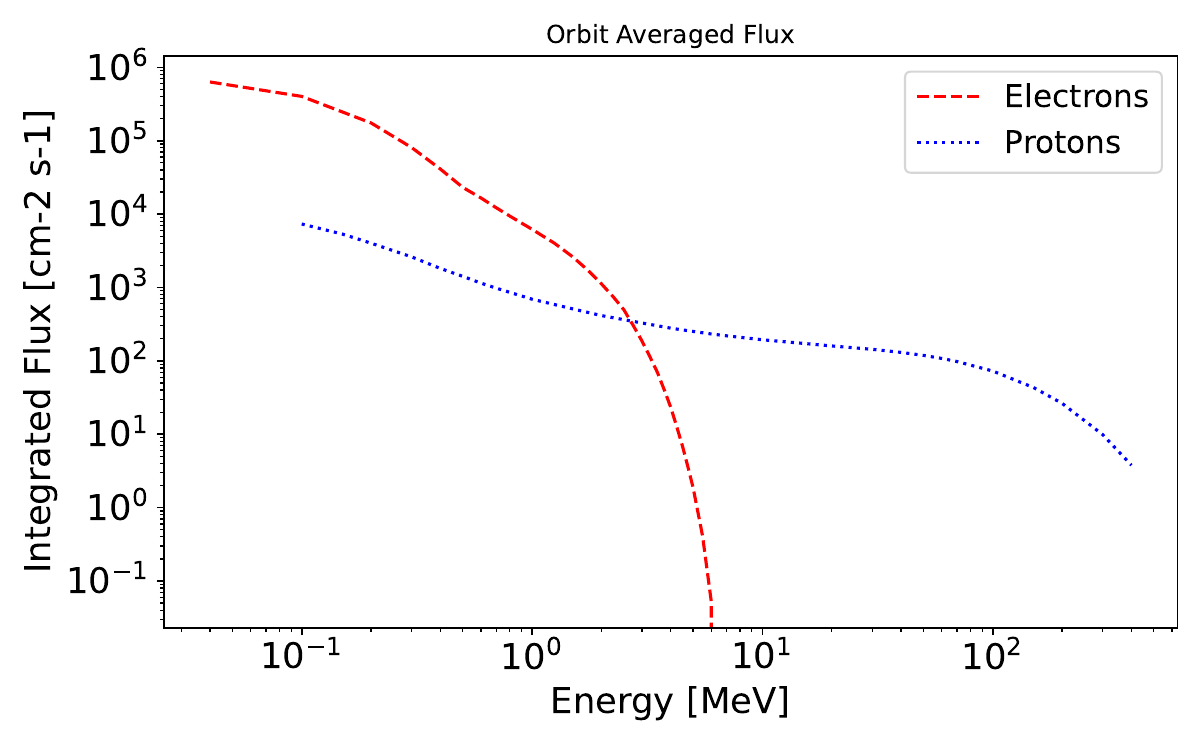}
\includegraphics[width=8.5cm]{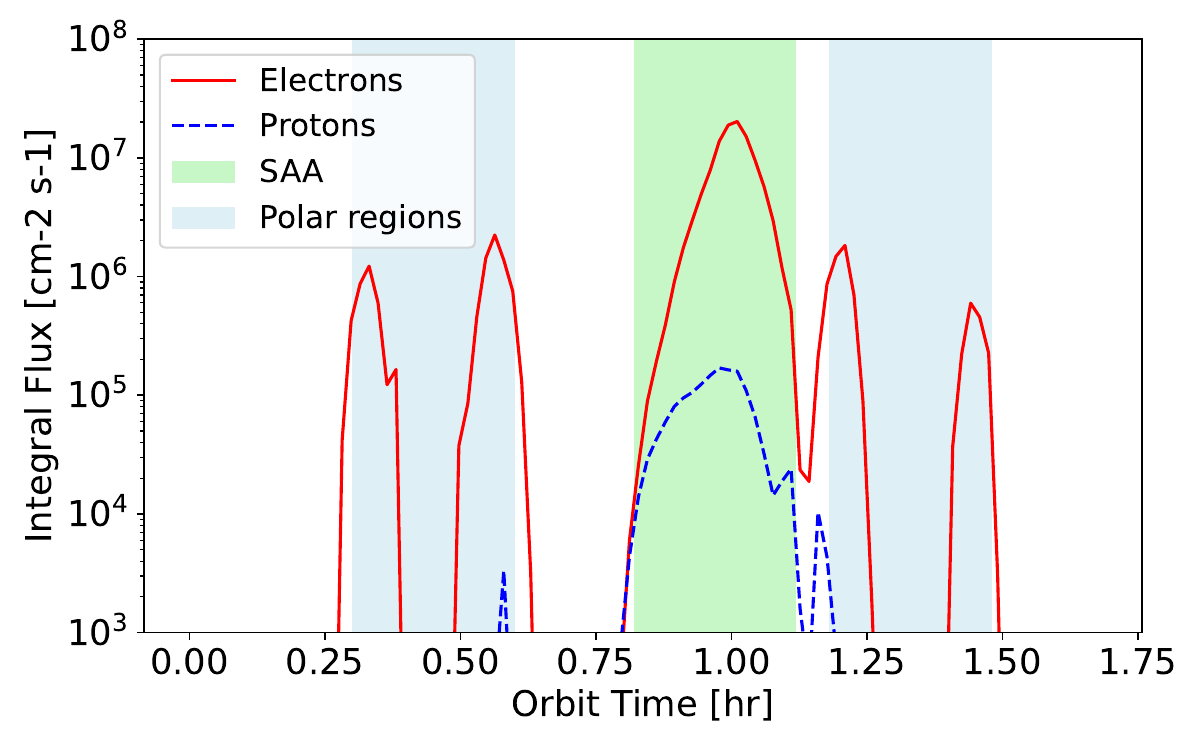}

\caption{\label{fig:rad_ev} (Top) Energy distribution of a 15-orbit-averaged radiation dosage consisting of electrons and protons at the S5p orbit. The average is over a length of 15 orbits with passages through the SAA on both the day- and night-side of an orbit. (Bottom) Time distribution of electrons and protons integral flux over an example orbit passing through the day-side of the SAA.}
\end{figure}

\subsection{Radiation impact expectation}
Impact from charged particles (predominantly protons and electrons) are the most likely source of damage within the HgCdTe substrate of the detector, potentially causing pixels to become inoperable.
The radiation dosage to which the whole S5p platform is exposed can be modelled using the Space Environment Information System \citep[SPENVIS, see][]{Heynderickx02, Heynderickx04, Kruglanski10} \footnote{SPENVIS is an engineering Operational Software of the European Space Agency maintained and operated by BIRA-IASB. BIRA-IASB is a governmental research centre depending on the Belgian Federal Science Policy. The tool used can be found at https://www.spenvis.oma.be/.}.
Given the S5p heliosynchronous orbit, and assuming AP-8 and AE-8 models for the proton and electron fluxes during solar minima, the orbit-averaged flux as a function of energy in mega-electronvolt (MeV) is given in Figure \ref{fig:rad_ev}. By number, the radiation dosage is dominated by low-energy electrons, with the exposure to higher energy protons being rare, but not negligible. 
In addition, Figure \ref{fig:rad_ev} shows the integral flux of electrons and protons during a typical S5p orbit that crosses the center of the SAA. Virtually all protons and the bulk of the electrons (90$\%$) of the orbit-averaged radiation dosage are encountered in the SAA. The remainder is seen during the crossings of a ring around the Earth's poles.
The typical radiation dosage during solar maxima is a factor 3 higher for energies below 1 MeV, and similar at higher energies. Given the temporal variation of Solar Cycle 24 \citep[e.g.][]{Zhang21}, S5P has flown predominantly during a solar minimum. 
Note that due to the shielding afforded by the casing around the detector, the number of electrons and protons reaching the detector is lower.

\subsection{Annealing}
For semiconductor material such as HgCdTe, the quality is not only determined by the properties of the material, but also its passivation process \citep{Mangin19}.  During long-term use, several processes can cause small defects to accumulate, increasing dark current and/or noise, thus reducing the performance over time \citep[e.g.][]{He15}.  Annealing, i.e. heating and subsequent cooling of the material under controlled circumstances, of the HgCdTe layer has been shown to improve the crystal state of the layer \citep[e.g.][]{Wang20}. This removes the small defects and can be effectively used to improve reduced performance.

Little is known about possible improvements on the performance in space conditions, nor are statistics on such detectors performance abundant.However, these types of tests are typically done by heating to temperatures between 273 K or 500 K under ambient conditions (i.e. not in vacuum). Some procedures on HgCdTe have been carried out by warming these up and subsequently cooling down. For more information, we refer the reader to \citet{Korotcenkov23} and references therein.

\subsection{Detector performance}\label{sec:det_noise}
As seen from earlier publications \citep{vanKempen19, Ludewig20, vanKempen23}, the detector has performed within parameters since launch, and is continuing to do so. No freeze-out of water has been detected \citep{vanKempen19}. The SWIR detector has been kept at its operational temperature throughout the mission, except for instances in which orbital control manoeuvres needed to be executed.

\begin{figure}[tp]
\includegraphics[width=9cm]{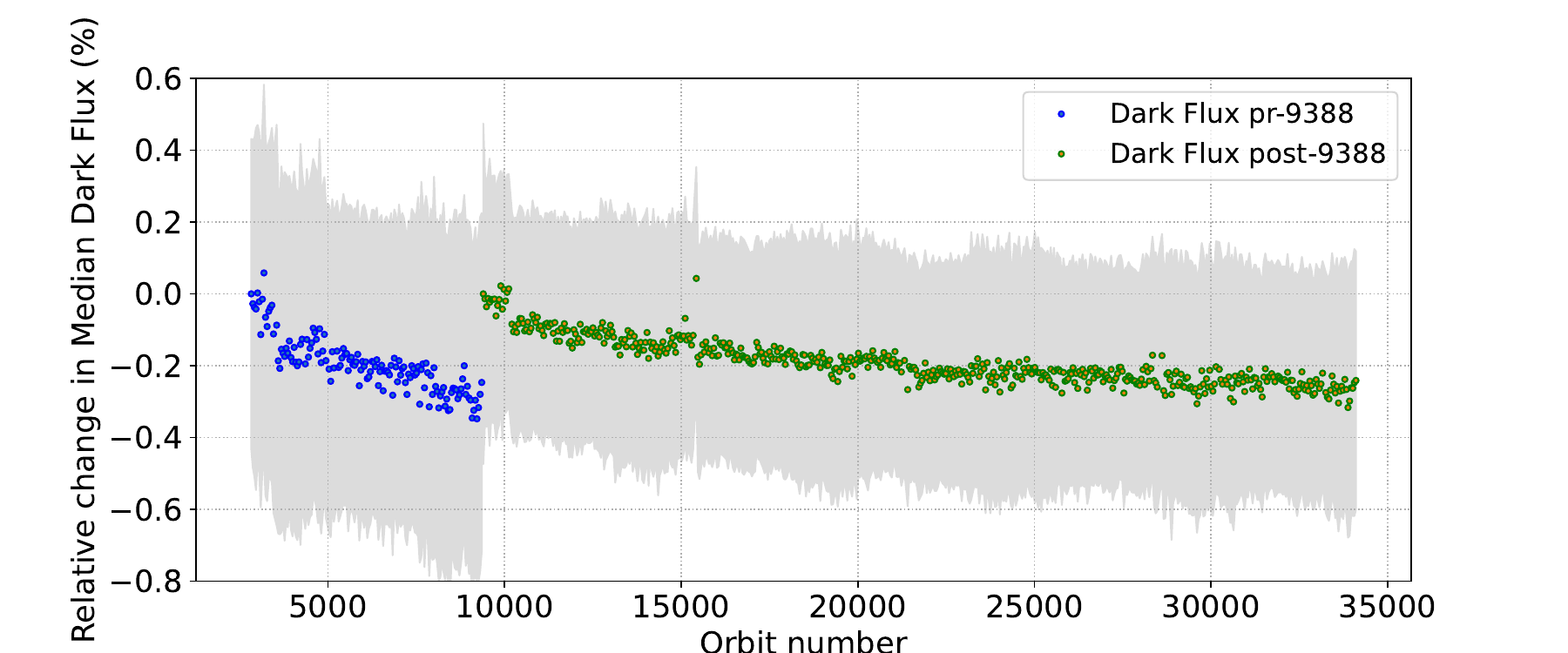}

\caption{\label{fig:dark} TROPOMI-SWIR detector signal produced by the dark current (i.e. the dark flux) as a median over the detector as a function of orbit number. Results are derived from radiance background measurements. The radiance background measurements copy the exposure times between 0.18 and 0.84 seconds of the radiance measurements. The dark flux is given as a relative percentage to the value at the reference orbit 2738.}
\end{figure}

The dark current of the detector has a median of 3738 +/- 12  e$^-$/s, a deviation of $<$0.2$\%$ \citep[ See][and Figure \ref{fig:dark}]{vanKempen19,vanKempen23}. There is a significant structure in the spatial distribution dark current, ranging between 3,000 to $\approx$ 5,000 e$^-$/s \citep{vanKempen19}. 

\begin{figure}[tp]
\includegraphics[width=9cm]{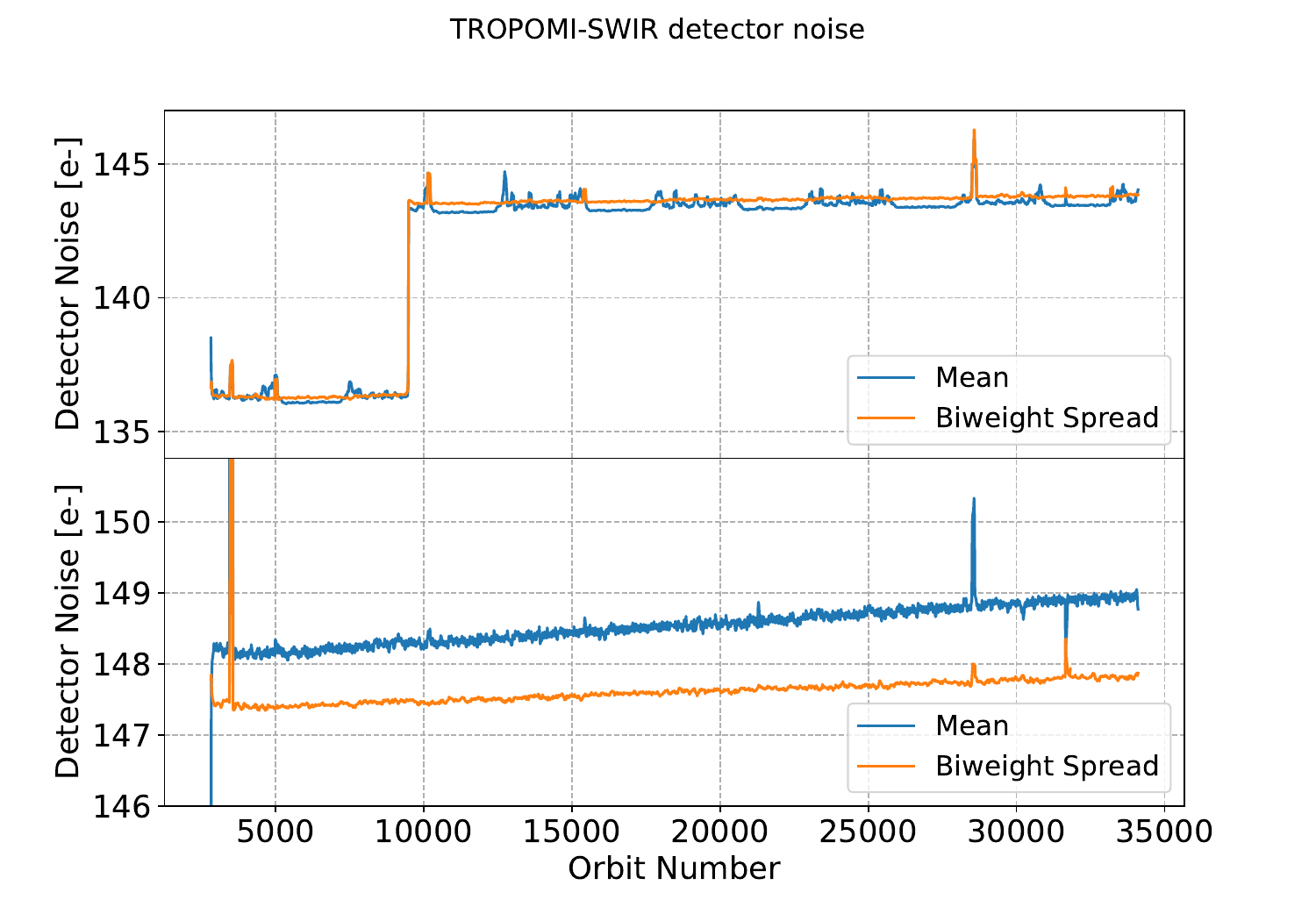}

\caption{\label{fig:noise} TROPOMI-SWIR detector noise as a median over the detector as a function of orbit number. Results are derived from radiance background measurements (top figure) and from dark measurements (bottom figure). The radiance background measurements copy the exposure times between 0.18 and 0.84 seconds of the radiance measurements,  while the dark measurements use a fixed exposure time of 1.08 seconds.}
\end{figure}

Readout noise was required to be better than 150 e$^-$. Early results provided a noise measured in-flight between 137 to 148 e$^-$\footnote{see https://www.sron.nl/tropomi-swir-monitoring/swir-noise} \citep[See ][ and Figure \ref{fig:noise}]{vanKempen19}. After five years this number has increased to 143 to 149 e$^-$. The noise is determined by either taking the standard deviation or a biweight spread of input data. The biweight spread is a one-step bi-weight scale estimator \citep[e.g.][]{Beers90}. This mathematical approach is less sensitive to extreme outliers (such as seen from cosmic ray impacts or broken pixels). For a perfectly gaussian distribution without outliers, the standard deviation and bi-weight spread are equal. 

The top panel in Fig. \ref{fig:noise} shows the results of both using the background radiance measurements for which the exposure times per scan are equal to that of the radiance measurements . The difference between the start of nominal operations and the current day is apparent, including a change in exposure time. The maximum exposure time of the radiance measurements was changed from 0.54 seconds (co-adding two consecutive measurements) to 0.84 seconds (no co-addition) to increase the spatial resolution. This was done in August 2019, orbit 9388. 
In the bottom panel, the noise has been derived from dedicated dark measurements during eclipse passages. Although more data is available, the measurements have a fixed exposure time of 1.08 seconds, resulting in significantly larger noise values. In this data, there is a clear small ($<$1 e$^-$) increase in noise over five years.  

In both derivations, transients are flagged and removed.  
Virtually all resulting noise values have been and remain below the requirement of 150 e$^-$. For the background radiance measurements increased noise levels during passages over the SAA can be seen in the mean. 
Interestingly, warm-ups of the detector caused by orbital control manoeuvres are not seen in either the dark flux or noise parameters.

\begin{figure*}[tp]
\includegraphics[width=17cm]{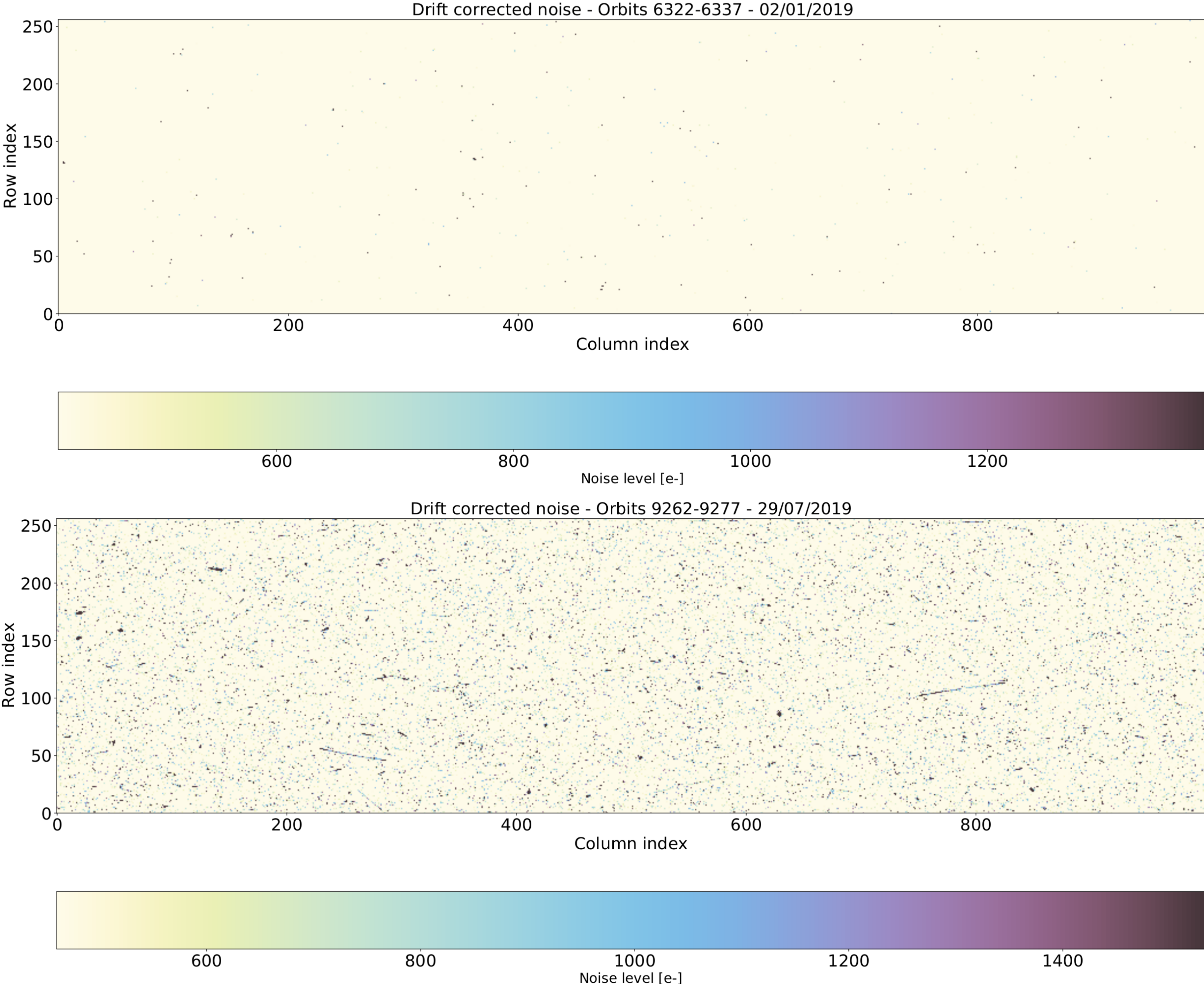}

\caption{\label{fig:radiation} Results for in-flight noise characteristics over the full SWIR detector. It shows the distinct effects passages through the SAA have on the detector. The noise parameters are derived from two distinct sets of dedicated calibration measurements. Each set consists of calibration data taken during 15 successive orbits (typically 7 or 8 usable measurements). At the top, all measurements are taken outside of the SAA over the northern hemisphere, while at the bottom the measurements are taken over the southern hemisphere, with three of the eight measurements close to or within the SAA region. The color scale for both images is three to ten times the median noise over the detector. The median noise is 138 electrons for the quiescent set and 153 electrons for the other set which includes the SAA measurements. The bottom image shows systematically higher excesses (in blue/purple) caused by cosmic ray impacts in the SAA.}
\end{figure*}

The immediate effects of particle impacts on the TROPOMI-SWIR detector can clearly be seen in the comparison of noise monitoring products, see Fig. \ref{fig:radiation}. Here, the noise performances, using the standard deviation discussed above, are shown on a scale starting from the median noise\footnote{138 e$^-$ and 153 e$^-$ for the first and second sets respectively. This difference originates from the change in observational parameters carried out in August 2019.  See \citet{vanKempen23}.}(yellow) to a maximum equalling the 99th percentile towards the value of most extreme outliers. The first set of measurements displays results related to the northern hemisphere, while the second set uses measurements in the southern hemisphere.
Each set consists of data from 15 consecutive orbits (orbits 6322 to 6337 and orbits 9262 to 9277). For the southern hemisphere set, three of the eight measurements cross the SAA. 
The difference in Fig. \ref{fig:radiation} is clear. There are a significant amount of features seen in the bottom plot from particle impacts (i.e., transients). Such temporary increase in noise are nearly absent in the measurements on the northern hemisphere. Identified transients do not imply nor exclude damage. Only subsequent lasting effects of transients are relevant. Similarly, the plots reveal that adopting the bi-weight spread results in noise performance independent of SAA crossings. \\

\section{Methodology for pixel quality determination}\label{sec:methods}

With the noise and dark current performance known, one must subsequently devise a method to quantify pixel quality.

\subsection{Pixel Quality}\label{subsec:flag}

The quality of detector pixels is determined and monitored through a number of tests. Each test result is expressed in a unit-less number between 0 and 1, with 0 corresponding to a completely broken pixel and 1 to a perfectly performing pixel. The overall quality of a pixel is determined by the minimum of all tests performed.  Pixels are considered to be acceptable for use in various retrieval algorithms if scores above 0.8 are achieved.


The set of algorithms\footnote{After the launch of S5p, some algorithms used in the tests and Calibration Key Data (CKD) were altered during several processor updates \citep[][ and the S5p document library, see https://sentinel.esa.int/web/sentinel/technical-guides/sentinel-5p/products-algorithms]{vanKempen23}. For consistency, this work uses the latest version of the algorithms deployed at the SRON SWIR monitoring system, applied to the data for the full mission. These are also referred to as the $'$beta-nominal$'$ version on the monitoring website of SRON. Older versions, including various operational algorithms used in instrument monitoring, are not included.} used to determine pixel quality before launch consisted of five tests. These five tests were used to identify bad and dead pixels and whether or not the detector met the required specifications.
The tests for the on-ground pixel quality assessment are:
\begin{enumerate}
\item \textit{Dark current}. The signal rate measured in the dark. It is divided by the median over the detector. If it is higher, the pixel quality is lower. 
\item \textit{Noise}. The random or semi-random variation in a a set of signals taken at the same exposure time. It includes read-out noise due to the hardware and shot noise due to the dark current and light. The shot noise due to light is assumed to behave normally (proportional to the square root of the dark-corrected signal), so signals taken in the dark are used. To correct for possible signal drift, the median over the detector is subtracted from each signal image before the noise is determined per pixel. That is divided by the median noise. If the noise is higher, the pixel quality is lower.
\item \textit{Radiance Responsivity}. Conversion factor from spectral radiance to signal rate due to external light. This test identifies pixels that are working properly but are not illuminated, at the detector edges. However, the edges are not included in the (ir)radiance data product. The responsivity can also find pixels that are not working, but the following test is meant for that. 
\item \textit{Quantum Efficiency}. The amount of charge generated in a pixel per incoming photon. If it is too low, the pixel quality is low. In practice, there are only 2 pixels with a quantum efficiency below 10$\%$ and those were already identified by their excessive dark current.
\item \textit{Memory Effect}. The amount of charge left on a pixel after a destructive readout that contributes to the signal seen by a subsequent readout. Using measurements with many different signal steps, the memory effect is characterized by the difference between the expected and the measured signal, divided by the difference between the measured signal and the previous signal. To be able to compare pixels, the absolute difference of the result with the median over all pixels is divided by the bi-weight spread over all pixels. This means if the memory effect deviates more from the median value, the pixel quality is lower.
\end{enumerate}
All tests were done on a per-pixel basis, with each test producing the unitless value between 0 and 1 described above. This is done by scaling the result using
\begin{equation}
f_{test} = (S_{test} - r_{test}) / (S_{test} - 1)
\end{equation}
followed by clipping to the range 0 to 1. Here, $S_{test}$ is the scaling parameter for the test, and $r_{test}$ the result of the test itself. $S_{test}$ can be used to provide weighting between tests, with lower values for $S_{test}$ corresponding to more important effects for lower quality.
The outcomes of each of the five tests ($f_{test}$) are combined to produce an overall quality index ($f_{total}$) between 0 and 1, derived by taking the minimum value among the five $f_{test}$ values. A possibility exist to manually provide a value below the final value. The final pixel quality according to this detector pixel quality flagging (DPQF) algorithm is $f_\mathrm{DPQF}$.\\

In flight, the tests on radiance responsivity, quantum efficiency and memory effects are not repeated; these turned out either not to be relevant or operationally difficult to even impossible to perform. 

A new test was added that identifies pixels that possess an acceptable value for the noise in one measurement, but have shown large variations in their noise performance over an extended period. The difference is taken between the 95$\%$ percentile and 5$\%$ percentile of the noise values and divided by the median noise value. On average this is 0.2, but it can change over time. Hence, the noise variation of a given pixel is divided by the median over all pixels. 

The dark current and noise parameters are derived using in-flight calibration measurements taken at the eclipse side of TROPOMI orbits. These measurements are done when the TROPOMI aperture is closed using the folding mirror mechanism. This is described in \citet{vanKempen19} and \citet{Ludewig20}. Noise variation is derived by comparing the measured noise level of a pixel in the noise test to previously obtained results for the noise. 

The dark current parameter $r_\mathrm{dark}$ is given by the dark current of a pixel divided by the median of the dark current over the detector. Its scale parameter $S_{dark}$ is set to 11. This value will produce a value of 0.8 when the dark current is 3 times the median. \\
The noise flag parameter $r_{noise}$ is given by the noise of a pixel divided by the median of the noise over the detector. The scale parameter $S_{noise}$ also equals 11. \\
The noise variation $r_{var}$ is subsequently calculated over the last 45 orbits. The scale parameter $S_{var}$ is again set to 11. As seen in Fig. \ref{fig:radiation}, taking either less orbits or a lower scale parameter will result in differentiation between subsequent tests due to the inclusion of widely varying number of passages through the SAA, where most detector impacts take place. \\
The final value for the quality index $f_{total}$  is called $f_{DPQF}$ in-flight, and is used to label a pixel as either $'$good$'$ ($f_{DPQF}$ value between 1 and 0.8), $'$bad$'$ ($f_{DPQF}$ value between 0.8 and 0.1) or $'$dead$'$ ($f_{DPQF}$ value below 0.1). The terminology is subjective as dead pixels may still show a signal response. It is recommended to only use pixels with a $f_{DPQF}$ value above 0.8.

\begin{figure}[tp]
\includegraphics[width=8cm]{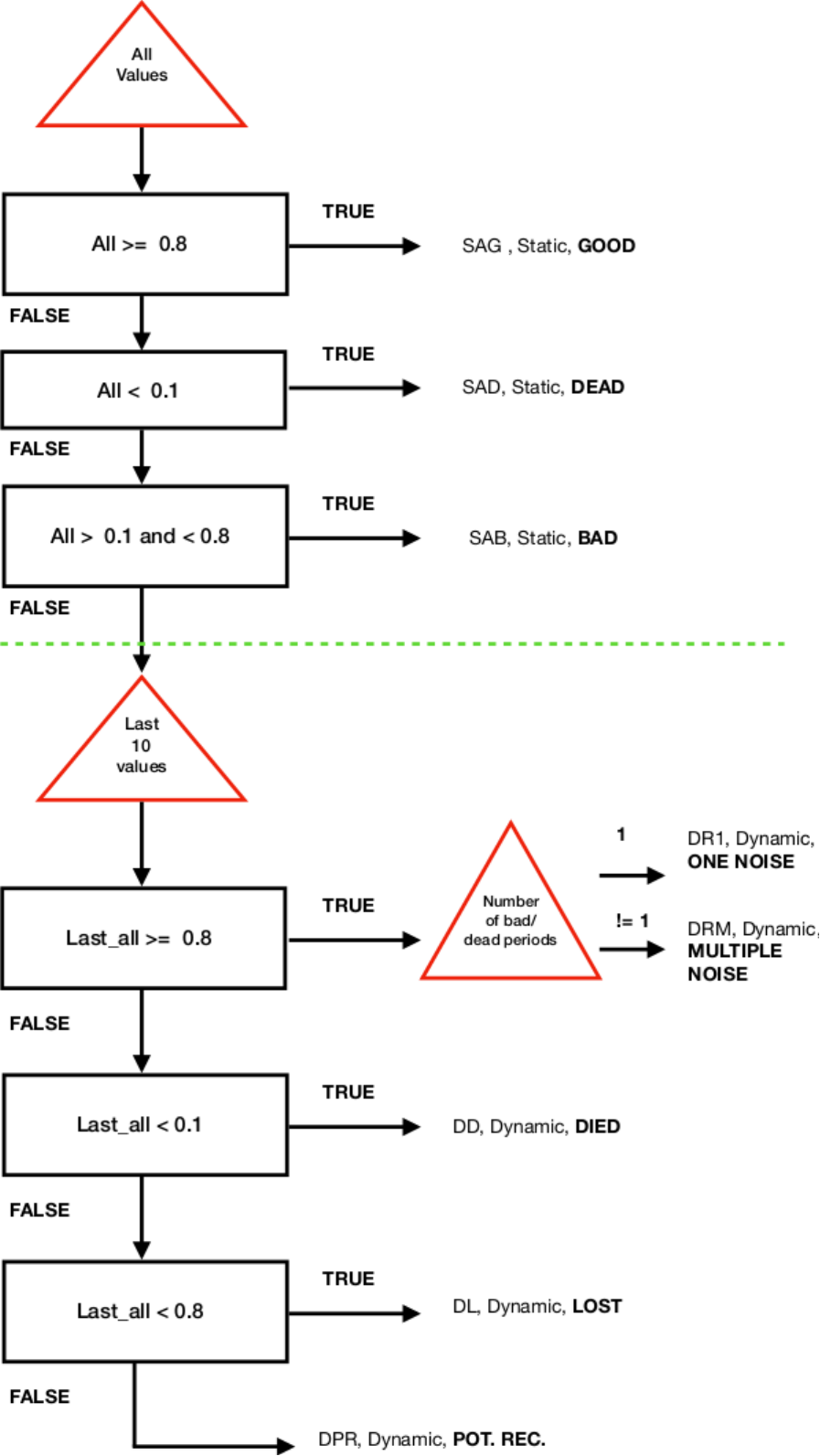}
\caption{\label{fig:flow} Flow diagram of the labelling of pixels.}
\end{figure}

%


\subsection{Quality Categories}\label{subsec:categories}
Detector degradation, either due to degradation in the material of individual pixels or adverse effects in the read-out electronics, is thus expected to be detectable in two ways. First, a sudden (i.e. from one orbit to the next) increase of the noise, noise variation and/or dark current (i.e. a drop in $f_{DPQF}$), or, second, in a significant drop in transmission over parts of, or the entire, illuminated area of the detector. Transmission loss is monitored by the response to the on-board light sources and solar irradiance measurements and originates in the degradation of the optical path, not the detector. It are considered to be beyond the scope of this paper. For recent results we refer the reader to \citet{vanKempen23} or the TROPOMI monitoring sites: \url{http://www.sron.nl/tropomi-swir-monitoring} and \url{http://mps.tropomi.eu}. \\

To structurally investigate lasting effects of detector degradation, not only must one consider the performance of a pixel at the time of inspection, but also changes over the full lifetime of TROPOMI-SWIR, starting at orbit 2756. This orbit itself is part of the set of reference orbits. These contain the first calibration measurements carried out during nominal operations. Measurements are averaged over sets of 45 orbits due to repeatability of the closing of the FMM \citep{Ludewig20}. 

Categorization is done for each interval. Thus, at each interval the historic performance of a pixel up to that interval is considered for the categorization. An overview of the categories, as well as the order in which the categorization is done, is illustrated by a decision tree in Fig. \ref{fig:flow} and described below. 

The categories themselves are also grouped in three super-categories for clarity. The first is the $'$Static$'$ (or $'$S$'$) super-category. This contains all pixels that have not changed labelling at any point in time since the reference orbit. Static contains three categories: i) SAG: pixels that have always been labelled as good, ii) SAB: pixels that have always been labelled bad, and iii) SAD: pixels that have always been labelled as dead. By definition the number of pixels in these categories can never increase.

The next super-category is $'$Dynamic$'$. Pixels in this category show one or more changes in its history up to the last measurement. In this category a change in labelling occurred. If a pixel has been labelled as good at the measurement time, they are either labelled as $'$DR1$'$ or $'$DRM$'$. The $'$DR1$'$ category contains pixels in which only a single period of lower labelling (i.e. non-good) took place; $'$DRM$'$ is used if multiple such periods exist. If a pixel is labelled as bad or dead, but was labelled good in the last month, a pixel will be categorized as $'$DPR$'$. This indicates a potential recovery for the pixel. 

Last but not least, pixels that have been labelled as bad or dead over a period longer than the last 30 days, are put in the last super-category $'$Unrecoverable$'$. This includes categories of $'$DL$'$, for pixels labelled as bad over the last full month, and $'$DD$'$, for pixels labelled as dead for a full month. These are unlikely to recover. The choice for the period of one month between the Dynamic and Unrecoverable super-categories originated from an initial expectation for the length of time when a pixel can or cannot be recovered. This initial assessment was done during the five-month E1 commissioning period \citep{vanKempen19}.  \\

\section{Results}\label{sec:inflight}

\subsection{Reference orbit}\label{subsec:reference}

After completion of the instrument commissioning \citep[For results on this period, see][]{vanKempen19,Ludewig20}, orbit 2756 is set as the first of the set of 45 reference orbits. It was executed on April 28th 2018. 

Table \ref{tab:init} gives the distribution of the pixel labels for the on-ground results, the reference (i.e., the first) entry in the monitoring system, as well as the a recent orbit (27712). 
There is a significant difference between the on-ground results and the reference results. For the reference, taken at the end of commissioning, over 99$\%$ of the pixels are labelled as good, with 0.1$\%$ labelled as dead, significantly better than the on-ground results, where just 98.5 percent was labelled as good. The amount of dead pixels has been reduced by a factor 5, while the amount of bad pixels was reduced by 27$\%$. Both distributions are evenly distributed over the detector.  
The origin of the observed improvement in the pixel quality has its origins in various effects. It is known that the thermal relaxation time of several components within the SWIR module is very long, ranging from days to weeks. The immersed grating in particular has a relaxation time that was measured in months. While in-flight the system was in relative thermal equilibrium for nearly six months at the time of the reference, tests on-ground were done with at most a few days to a week of thermal relaxation time. The surface used for the background measurements on-ground similarly may have not been at the desired level of thermal control. In addition, the exposure times and number of points used to derive the quality parameters on-ground differ from the ones used in-flight.

\begin{figure}[!tp]

\includegraphics[width=8.5cm]{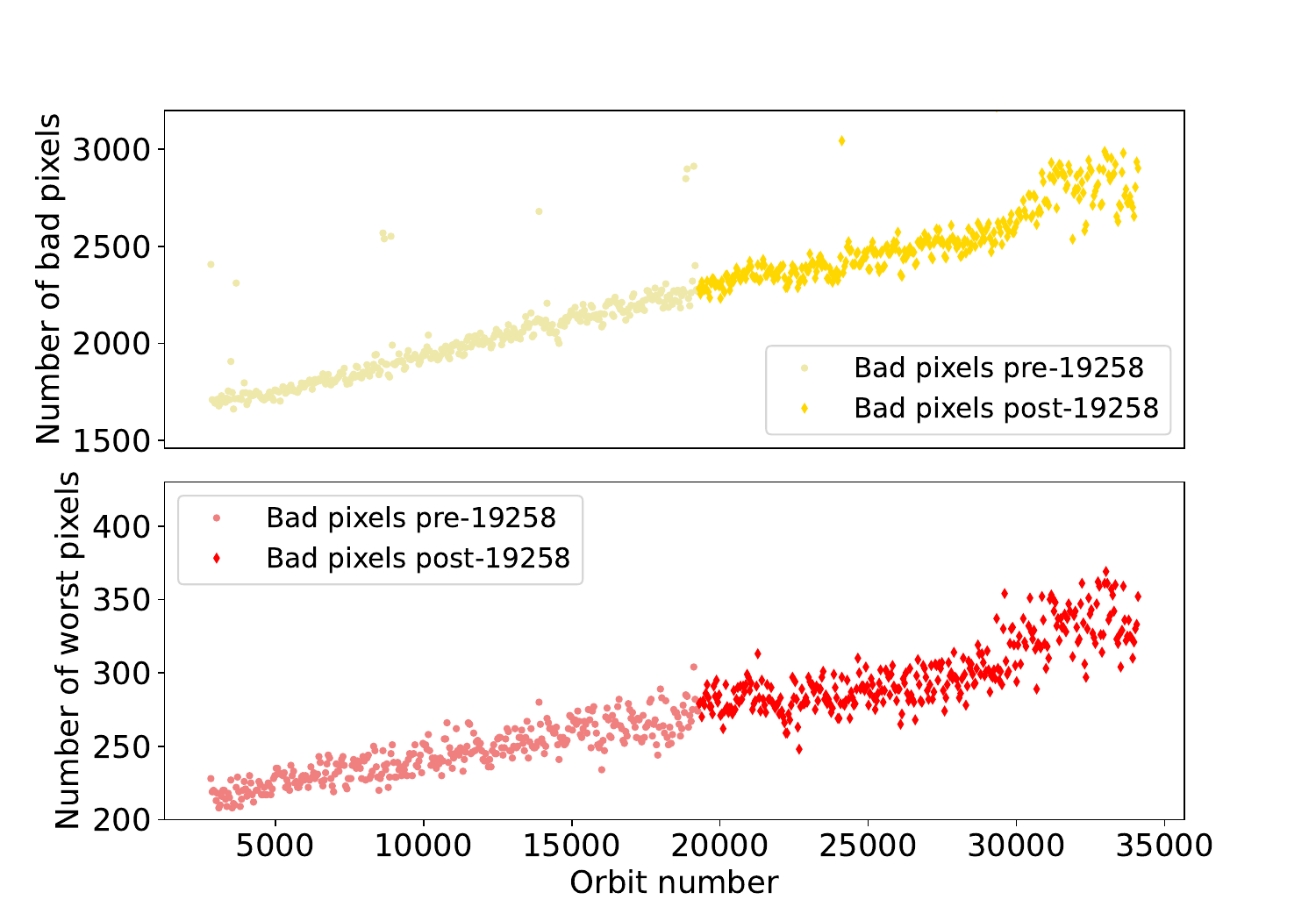}

\caption{\label{fig:qua} The number of detector pixels identified as bad (0.1 $<$ quality $<$ 0.8) and dead (quality $<$0.1) over the illuminated part of the detector from the start of nominal operations (April 28th 2018) to the current day. Orbit number 25000 occurred during November 2022. The updates in exposure time (orbit 9388) and calibration key data (orbit 19358) are indicated with a change in symbols or colors.}
\end{figure}

\begin{figure*}[!th]
\includegraphics[width=17cm]{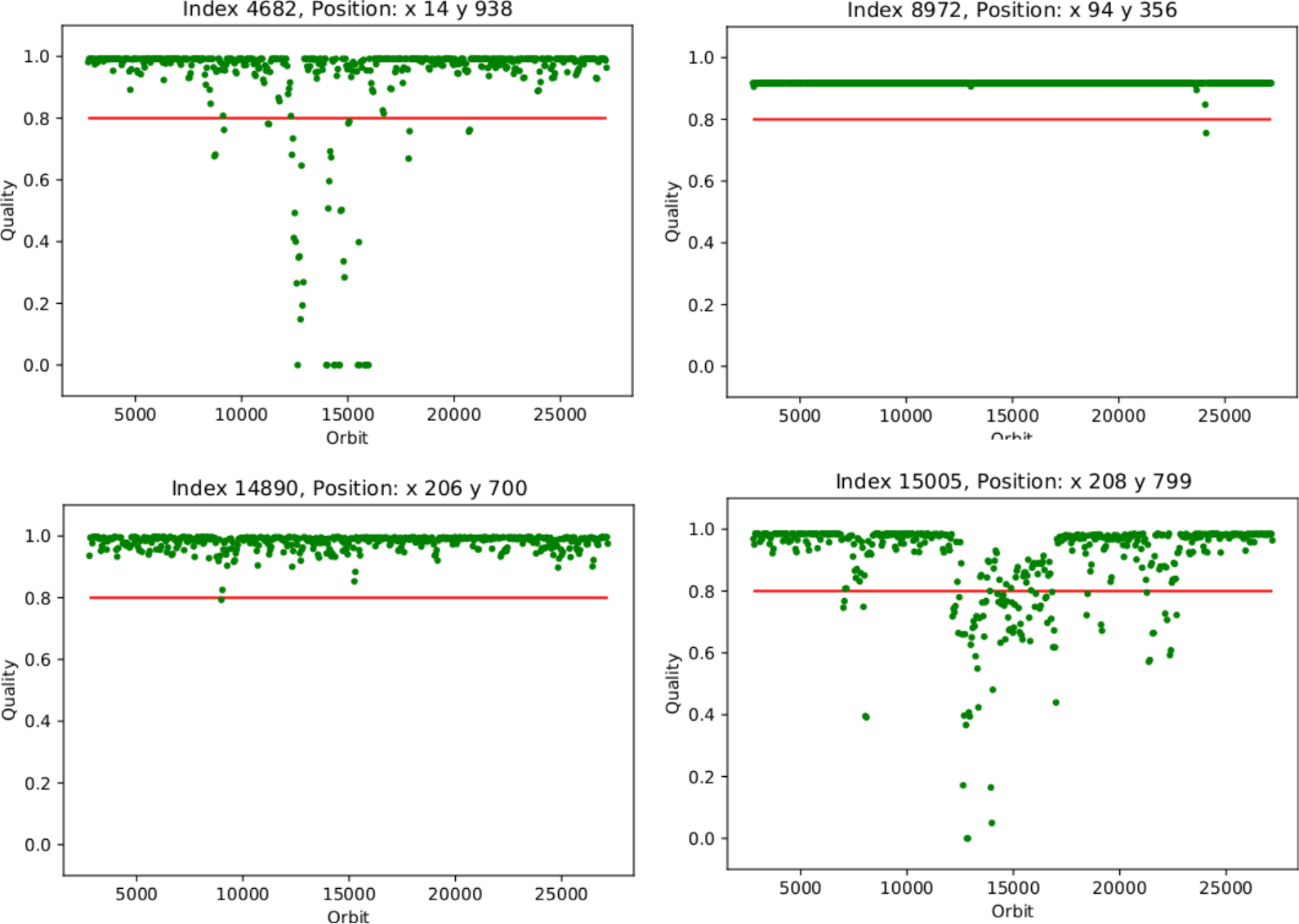}

\caption{\label{fig:indiv} Examples of the time series of individual pixel labelled in the  dynamic super-category. These include a variety of occurrences where pixels are labelled bad or even dead in the case of 4682 (top left). It includes a pixel with at some point a quality with a value of 0.0 with a subsequent recovery (I.e., pixel 4682), pixels with only a single occurrence in a different category (pixel 8972, top right, and 14890, bottom left), and a pixel with many occurrences (pixel 15005, bottom right). }
\end{figure*}

\subsection{In-flight detector quality monitoring}\label{subsec:inflight_mon}

Figures \ref{fig:dark} and \ref{fig:noise} show the median dark current and the median noise over time. Figure \ref{fig:qua} subsequently shows the amount of pixels labelled as either bad or dead from these results and the noise variation test.
Both the number of bad and dead pixels appear to have a linear trend upwards, albeit with significant scatter. Deviations from a perfect linear trend are indicative of local changes at other time scales. I.e., pixels identified as bad or dead at a previous interval must have changed labels (i.e. recovered), and thus categories. If not, there cannot appear scatter. This also reveals that pixels may able to both degrade further (change from bad to dead). Inspection of the time series of individual pixels confirms this. 
Figure \ref{fig:indiv} shows the pixel quality value of four randomly selected pixels as a function of orbit (i.e.,  time) with at least one point in time in which the quality is below 0.8.

\begin{table}[!th]
\begin{tabular}{l l l}
\hline \hline
Category & Percentage & Number of pixels \\
\hline
\multicolumn{3}{c}{On-ground results}\\
\hline
Good & 98.48 & 246567 \\
Bad & 1.00 & 2511 \\
Dead & 0.52 & 1290 \\
\hline
\multicolumn{3}{c}{Reference}\\
\hline
Good & 99.17 & 248302 \\
Bad & 0.73 & 1817 \\
Dead & 0.10 & 249 \\
\hline
\multicolumn{3}{c}{Orbit 27172}\\
\hline
Good & 98.77 & 247276 \\
Bad & 1.09 & 2741 \\
Dead & 0.14 & 351 \\
\hline

\end{tabular}
\caption{\label{tab:init} Initial distribution on-ground (top), at the reference orbit 2756 (middle) and at the current day at orbit 27172 (bottom) of the number of flagged pixels and percentage of the total. Edge columns are not considered.}
\end{table}

\begin{figure}[!th]
\includegraphics[width=8.5cm]{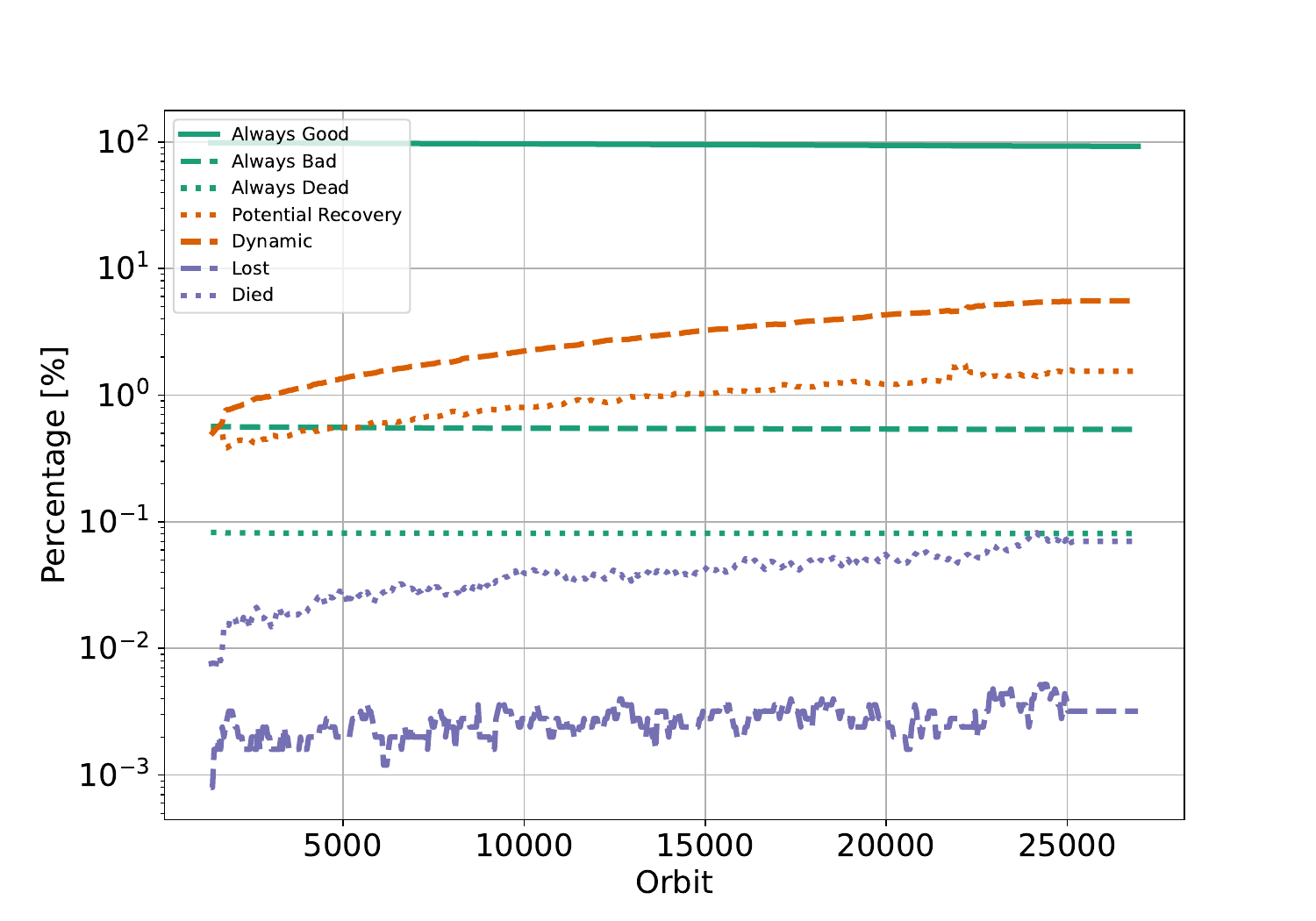}

\caption{\label{fig:evolution} Evolution over time of pixels in the individual categories defined in Figure \ref{fig:flow} as percentage of the total number of detector pixels. }
\end{figure}

Figure \ref{fig:evolution} in turn shows the evolution of all pixels over time as the percentage of pixels within individual categories, starting from orbit 2801 (i.e. the first entry that can be compared to the reference). The reference does not include any dynamic points and has been omitted from the figure. By definition, static categories will drop over time.
The vast majority of pixels ($>$95$\%$) has at all times remained always good.

\newpage

\section{Analysis}\label{sec:analysis}
\subsection{Degradation occurrences}\label{sec:degrad}
Using the results on categorization above, one can perform a statistical analysis to look at possible origins. Multiple or long-lasting degradations indicate permanent damage, while single occurences correlate to more superficial damage. The analysis revolves around two relevant parameters for the degradation of any individual detector pixel: a) The number of degradations seen for a pixel over the nominal operations time, and b) the time length of individual degradations. The smallest unit of time is the interval measurements; i.e. 45 orbits, or three days. Only pixels with at least one degradation in one time unit are considered (i.e., all pixels in the static super-category are omitted). 

\begin{figure}[tp]
\includegraphics[width=8.5cm]{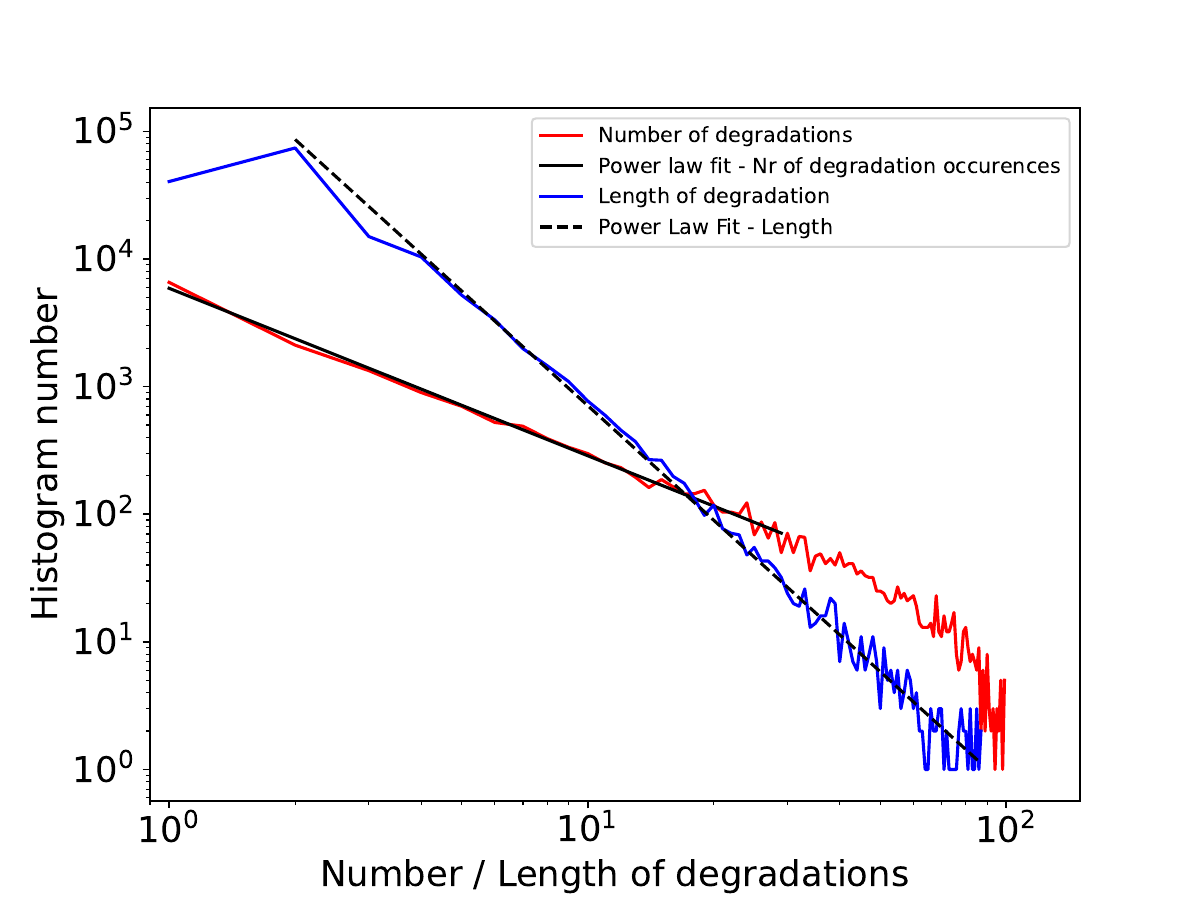}

\caption{\label{fig:power_law} Power Law distribution for the length of each degradation (blue) and the number of times a pixel degrades (red).}
\end{figure}

Figure \ref{fig:power_law} gives the probability density functions for the number of degradations seen over the detector (red), as well as the distribution of the time length of each degradation occurrence (blue).
Both distributions are fitted with a power law that is limited in range (solid and dashed black lines). 
These fitted power law distributions (represented as $f(x) = ax^{-k}$ for $x_{max} > x > x_{min}$) are classified as Type I Pareto distributions \citep{Arnold83, Newman05}. The slopes given by the exponents $k$ can be derived by using a least squares regression analysis; 1.3 for the number of degradations and 3.0 for the length of degradations.  The parameter $x_{min}$ is estimated to be 2 for the time length of a degradation and 1 for the number of degradations in each pixel.

The deviations from the power law fit in the two extremes deserve inspection. First, the number of degradation occurrences with a time length of 1 is severely  under-produced. Given the slope of the fitted distribution of 3 for the length of the degradation, about 300,000 degradations with a time length of 1 should have been seen. In reality only 40,000 ($\sim$13$\%$) were observed. Similarly, pixels with very many occurrences (excess of 30 occurrences) are under-represented. 

The origin of these observed deviations can in part be explained due to the data being incomplete. Technically, Pareto distributions are not constrained by $x_{max}$, but continue onwards in an infinite domain. Given the limits on time in the form of the total length of nominal operations, it isn't surprising both distributions thus deviate from the fitted Pareto distribution. There hasn't been enough time. For example, 100 degradations of length 2 would require a minimum sampling of 400 over the full lifetime. This is only marginally better than the 500 points available. With the relatively low number of pixels showing degradations at all, the probability of no pixel showing this behaviour is large.
Even with five years in orbit, there has not been enough time for pixels to fill the upper end of the distribution for the number of degradation occurrences. Given enough time, the distribution for the number of degradation will be filled out.

The under-representation of degradations of length equal to one (i.e., 45 orbits) does not originate from limited sampling and  finite size of the dataset. The origin of this deviation must be physical in nature, likely in the substrate of the pixel. A minimum time-scale of at least two units of time lengths is typical for an occurrence. This equals 90 orbits or 6 days. \\
 
The mean of a Pareto type power law distribution (given by $(k-1/k-2)\times x_{min}$) only exists for $k > 2$ \citep{Arnold83, Newman05}; An expected mean for the number of degradations is thus not defined. The mean length of a degradation in time is 4 time units (assuming $x_{min} = 2$, thus equalling 180 orbits or 12 days). 

The median of both power law distributions can be derived, given by $2^{(1/(k-1))} * x_{min}$. For the length of time of degradations, a median value of 2.8 time units, equalling 127 orbits, is found. The number of degradations per pixel has a median value of 10 times, assuming $x_{min}$ of 1. For both distributions, the variance cannot be meaningfully defined \citep{Newman05}.  \\

\begin{figure}[tp]

\includegraphics[width=8cm]{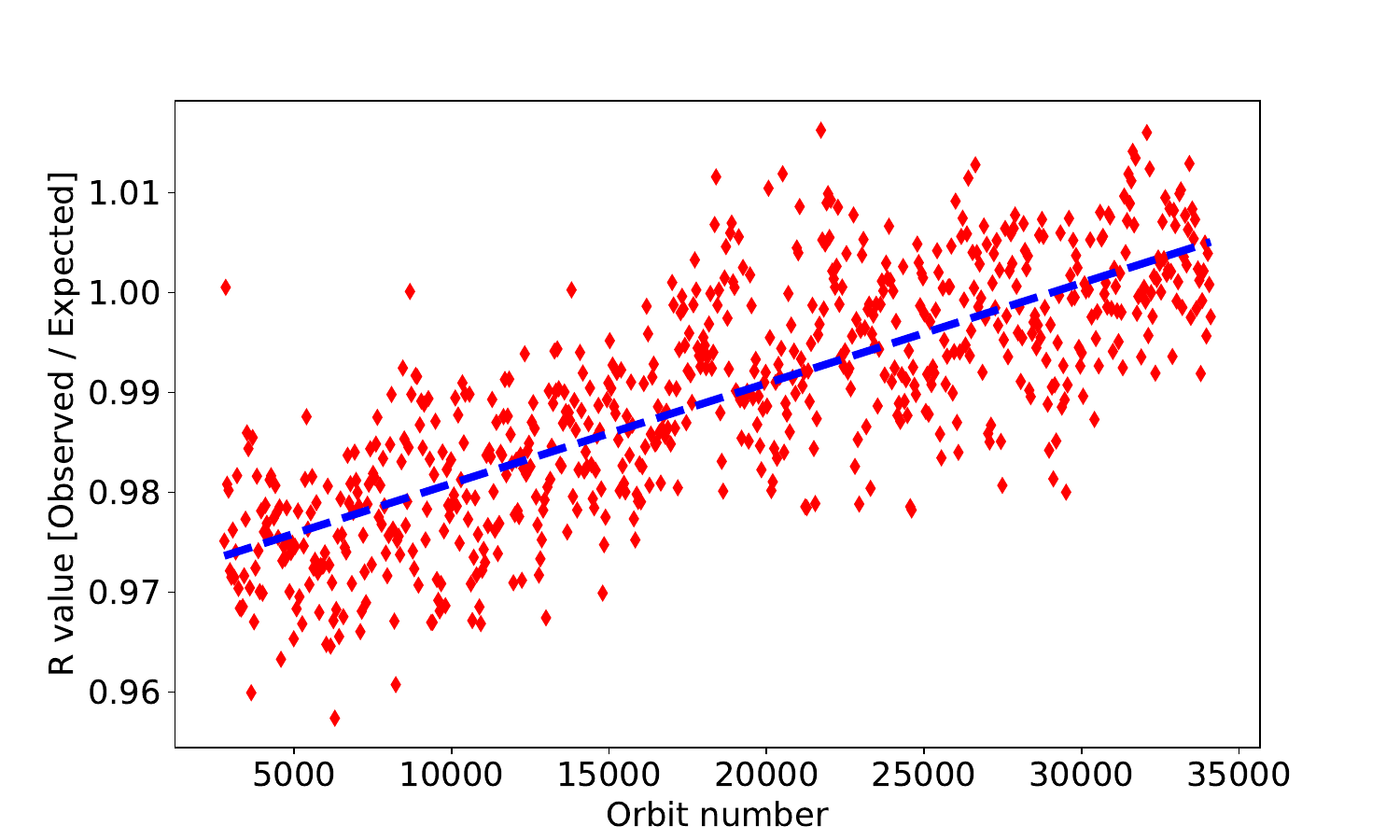}
\includegraphics[width=8cm]{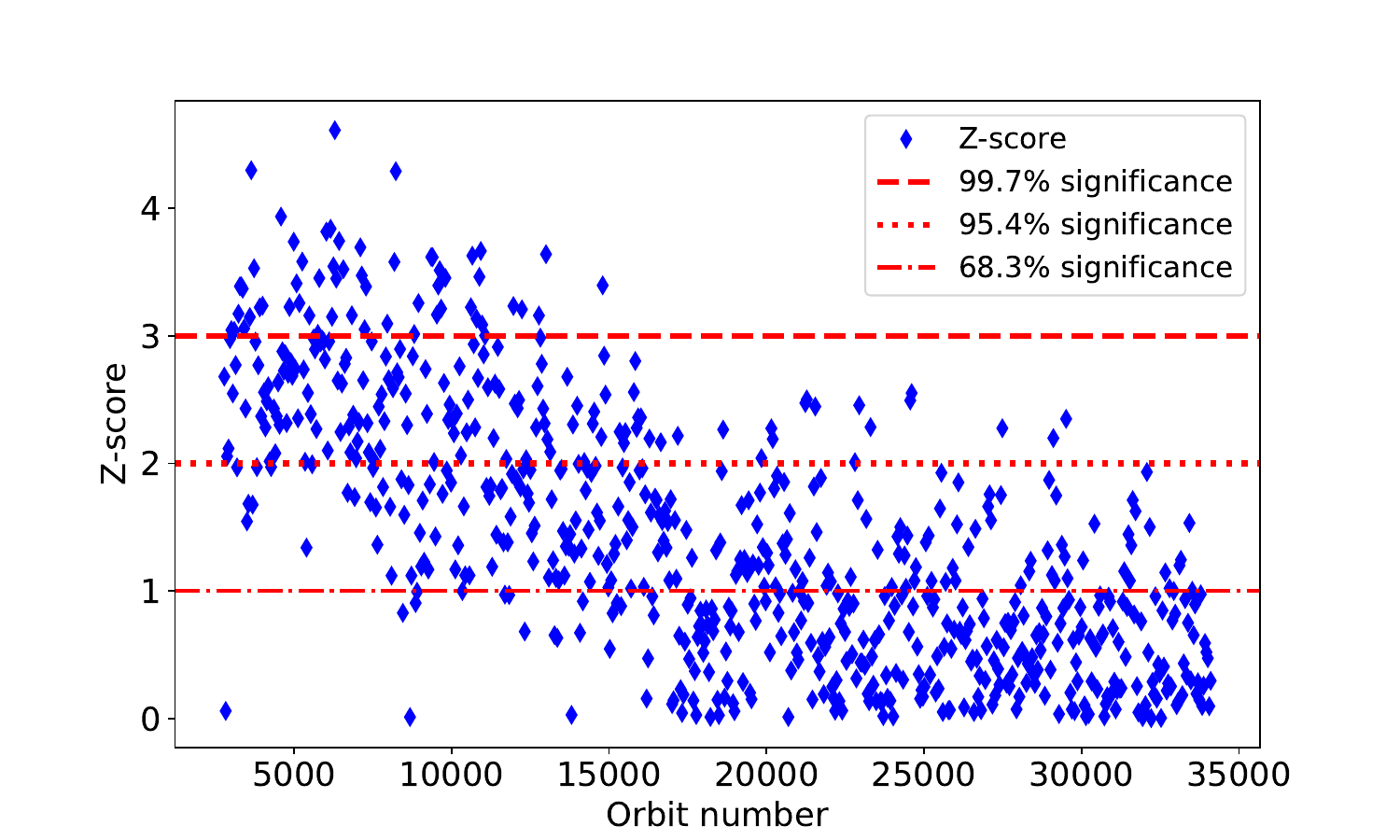}

\caption{\label{fig:cluster} (Top) Evolution of clustering parameter $R$ as a function of orbit. (Bottom) Associated Z-score compared to a random distribution.}
\end{figure} 

\subsection{Clustering}\label{subsec:location}

The spatial distribution of bad and dead pixels can reveal information on the origin of the physical impacts onto the detector that cause degradation occurrences. Particles with  higher energies are expected to damage clusters of pixels as it deposits charge that will bleed into nearby pixels. If the distribution of bad and dead pixels shows more regularity than a completely random distribution, this can indicate issues with the amplifiers and/or readout electronics, as these are not connected on an individual level, but to larger structures (e.g., rows and column). A completely random distribution shows that the detector is affected by cosmic rays with relatively low energies (i.e. the charge deposition occurs solely on one pixel), with the degradation primarily affecting the semi-conductor material of the pixel that was hit.\\

A significant number of statistical methods exists to determine the level of clustering, randomness and regularity \citep[e.g.][ and many others]{Clark79, Ripley81, Ahuja82}. The most useful approach for our case is described in \citet{Dry12}. Clustering is treated as a combination of the well-known travelling salesman problem and the minimum spanning tree problem. The approach is based on the methods independently derived in \citet{Hertz09} and \citet{Clark54} and summarized below as applied on the SWIR detector.

For a distribution of $n$ faulty pixels within an area $A$, the likelihood of finding a number of defective pixels within a given distance when pixels follow a completely random distribution is described by a Poisson distribution. The probability density function $p$ for pixels with respect to the distance to the nearest neighbour  $d$ is then given by 
\begin{equation}
p(d) = 2 \pi d \delta ~ e^{-\pi d^2 \delta}
\end{equation}
Here, $\delta$ is the point density (i.e., the mean number of points per unit area), equal to $n$ over $A$. 

The expected average distance of a collection of randomly distributed points to their respective nearest neighbour ($r_E$) is 
\begin{equation}
r_E = \sqrt{(A / n)} / 2
\end{equation}
with an associated standard error equal to
\begin{equation}
SE_E = \sqrt{(r_E / n)}
\end{equation}
The terms $r_E$ and $SE_E$ quantify the value for a completely random distribution of bad and dead pixels. 

To measure clustering and/or regularity, a measurable counterpart to this term is needed. If the distance of a point $i$ to another point $j$ is given by $r_{i,j}$, the observed average distance for the nearest neighbour  ($r_O$) is expressed as
\begin{equation}
r_O = \frac{1}{n} \sum_{i \neq j}^n (min(r_{i,j}))
\end{equation}
We assume the measurement error to be not applicable. 
The level of clustering, randomness or regularity is then expressed in term $R$ by dividing the observed distribution with the expected value for a Poisson distribution:
\begin{equation}
R = r_O / r_E
\end{equation}

A value of 1 for $R$ is a fully randomized (i.e. Poissonian) distribution of faulty pixels, while a value below 1 reveals clustering. A value above 1 shows more regular distribution. We refer the reader to the illustration in figure 2 of \citet{Dry12} for a clear visual representation.
One must consider the Z-score of the observed distribution $r_0$ to assure that the measured distribution is not a representation of an underlying Poissonian population.  The Z-score is expressed as:
\begin{equation}
Z_R = (|r_O - r_E |) /SE_E
\end{equation}
A high Z-score corresponds to an assurance $r_0$ and $r_E$ are not both representations from an underlying distribution. E.g., a Z-score of 2 would correspond to a 96.5$\%$ reliability the observed distribution is not drawn from a random distribution.

For the reference orbits near the start of nominal operation, a value of 0.9751 is found, with an associated Z-score of 2.68, which corresponds to a 99$\%$ certainty $r_O$ is not drawn from the same population as $r_E$. I.e., a distribution that is nearly random, but has a minor bias to clustering due to a few faulty regions.

Using the data obtained over the five years of TROPOMI-SWIR monitoring, Figure \ref{fig:cluster} subsequently plots the evolution of clustering parameter $R$ as a function of time (i.e., in the form of orbit number), as well as the Z-score associated with it. Over the lifetime of TROPOMI-SWIR, the original distribution has slowly evolved to a distribution with a complete random distribution.  The increase is determined to be 0.035 in $R$ between orbits 2782 and 27172, as well as a Z-score below 1 since orbit 17000. The variance around the mean trend of $R$ is 0.007 and does not appear to increase or decrease of time. We note that the original clusters, which are in the always dead category, still remain visible in the current distribution. 

From this, two conclusions can be drawn:
\begin{enumerate}
\item No new damage clusters have emerged from substrate damages due to high-energy cosmic ray impacts. 
\item A trend towards R = 1 is seen, indicative that degradation is related to a Poissonian distribution. From the Z-score it can be concluded the distribution of faulty pixels can be described by a completely random distribution since orbit $\sim$17000.
\end{enumerate} 

\section{Discussion}\label{sec:discussion}

\subsection{Permanently lost vs recoverable pixels}
In the initial categorization, a choice was made to set one month as the time differentiating the Unrecoverable and dynamic super-categories. I.e., it was assumed pixels that were either bad or dead for a full month would not be able to recover. This time length was estimated from the (limited) data available from the commissioning period from December 2017 to April 2018. From the statistical analysis, the distinction of one month has no physical origin. In fact, given the results in section \ref{sec:analysis} and Fig. \ref{fig:power_law}, there are no clear time scales to correctly capture the distinction between unrecoverable and recoverable. Figure \ref{fig:power_law} shows that recoveries have taken place up to at least $\sim$240 days, nearly eight months. This maximum recovery time likely will grow during the lifetime of TROPOMI. 
The Pareto power law distribution does provide a mean and median, but, by the nature of the distribution, these should not be interpreted as representations for a typical time-scale for a pixel recovery.  With the available data, all dead and bad pixels appear to have a potential recovery time, that can be months or even years. 
Only pixels that were identified on-ground as dead with a large dark current should be treated as permanently lost (e.g., the cluster of damaged pixels near column 450 and row 25.).  

\subsection{Lack of clustering}
One of the interesting results of the analysis seen in Fig. \ref{fig:cluster} is the fact that the distribution of bad and dead pixels remains nearly randomly distributed over the time range investigated. It was known that a few clusters had been damaged before launch, which agrees with the $R$ value just below 1 at the start. But the value of $R$ has remained very close to unity. This has two strong implications. First, individual impacts rarely, if ever, damage clusters of pixels. Most likely, higher energy particles able to cause clusters of damaged pixels are captured by the outer Van Allen radiation belt, with which S5p does not interact.
Similarly, the so-called $'$snowball$'$ features seen in JWST NIRSPEC and NIRCAM \citep{Birkmann22,Rieke23} are absent. These are never detected. Such events would even temporarily show up as a large cluster (e.g., the NIRCAM results show clusters of 50 by 50 pixels). But the R value remains close to unity at all times.
Third, it implies that in addition to the process that damages the detector being random in location, the recovery process is also best described by a randomly occurring process, not dependent on the location of the bad and dead pixel on the detector. 
This reinforces the assumption that the damage of bad and dead pixels occurs in the substrate and is not occurring in an electronic component. 

\subsection{Spontaneous pixel recovery}

The results presented in Section \ref{sec:degrad} reveal that most pixels degrade due to the continuous radiation impacts. This is evident from the random distribution. The recovery is on a finite time-scale, albeit with an undefined typical time. 
However, no active steps were taken to repair damaged pixels in the detector at any time during TROPOMI-SWIR operations (i.e. by a scheduled warm-up/annealing of the detector). With the observed recoveries and temperature control of the SWIR detector, it must thus be concluded that the detector must be spontaneously recover pixels by itself during operations. Annealing is a known property of HgCdTe detectors \citep{Lei15}, but requires heating and subsequent controlled cooling of the material. 

The observations of this effect in operation has significant impact on detector selection of future space missions in the infrared. The spontaneous pixel recover effect should be considered when selecting and characterizing HgCdTe detectors for such cryogenic infrared instruments, regardless of its scientific goal. For instance, it is encouraged to develop HgCdTe detectors with a cut-off wavelength of 1.8 $\mu$m, to replace the much more radiation sensitive InGaAs detectors commonly used up to 1.65 $\mu$m.

\section{Conclusions}\label{sec:conclusions}

In this paper, we have presented results for the in-flight performance of the TROPOMI-SWIR HgCdTe detector and observed damage due to hits of cosmic rays. The detector performance has been excellent, continuing the trend first seen in \citet{vanKempen19} and \citet{Ludewig20}. For the orbit of S5p, the radiation dosage to which the SWIR detector is exposed to is dominated by the low-energetic flux during passages of the South Atlantic Anomaly. These electron and proton impacts dominate the transients seen in the detector signals.
The summarized conclusions from the analysis presented are:
\begin{itemize}
\item The increase in inoperable (i.e., bad and dead) pixels is marginal. It has increased from 0.825 $\%$ at the start of the nominal operations to 1.235$\%$ measured in orbit 27172, an increase of 0.4$\%$. All fractions are considered negligible for scientific exploitation. 

\item Over 90$\%$ of dead and bad pixels are able to spontaneously recover, at the continous operational temperature of 140 K. The time-scales at which this occurs is described by a Pareto distribution with a mean recovery time of 12 days.

\item Pixels recovered from a degradation have a significantly higher chance to degrade again. The chance on re-degradation is also be described by a Pareto distribution with a median for the number of degradations of 10 over nearly five years.

\item The distribution of bad and dead pixels is random. I.e., there appear to be no clusters of damaged pixels created from impacts. The value $R$ for the amount of clustering, randomization and regularity has been derived, revealing a near-poissonian distribution.  

\end{itemize}

For future missions, the observed spontaneous recovery feature of the HgCdTe detector in TROPOMI-SWIR should be considered during mission design.

\ack
This paper contains Copernicus Sentinel data. This research is funded by the TROPOMI national program from the Netherlands Space Office (NSO).  \\

\def\newblock{\ }
\bibliographystyle{agsm}

\end{document}